\documentclass[12pt,a4paper]{article}
\usepackage{amssymb,amsmath,amscd,epsfig}
\title{Neutrino wave function and oscillation suppression.}
\author{
A.D. Dolgov\thanks{e-mail: dolgov@itep.ru}\hspace*{2mm}$^{\rm
a,b}$, O.V. Lychkovskiy, \thanks{e-mail:
lychkovskiy@mail.ru}\hspace*{2mm}$^{\rm c}$, A.A. Mamonov,
\thanks{e-mail: mamonov@dgap.mipt.ru}\hspace*{2mm}$^{\rm c}$, \\
L.B. Okun\thanks{e-mail: okun@itep.ru}\hspace*{2mm}$^{\rm
a}$,  and
M.G. Schepkin\thanks{e-mail: schepkin@itep.ru}\hspace*{2mm}$^{\rm
a}$\\[5mm]
${\rm ^a}$ {\small\it Institute of Theoretical and
Experimental Physics}\\ {\small\it 117218, B.Cheremushkinskaya 25,
Moscow, Russia}\\
${\rm ^b}$ {\small\it INFN, Ferrara 40100,
Italy} \\
${\rm ^c}$ {\small\it Moscow Physics and Technology
Institute}}

\date{}

\begin{document}
\newcommand{\be}{\begin{eqnarray}}
\newcommand{\ee}{\end{eqnarray}}
\newcommand{\bi}{\bibitem}
\newcommand{\paz}{p_A^{(0)}}
\newcommand{\plz}{p_l^{(0)}}
\newcommand{\tin}{t_{in}}
\newcommand{\vpB}{{\bf p}_B}
\newcommand{\vxA}{{\bf x}_A}
\newcommand{\vxa}{{\bf x}_a}
\newcommand{\vxB}{{\bf x}_B}
\newcommand{\rar}{\rightarrow}
\newcommand{\vx}{{\bf x}}
\newcommand{\vxc}{{\bf x}_C}
\newcommand{\vxf}{{\bf x}_f}
\newcommand{\vpd}{{\bf p}_D}
\newcommand{\vplp}{{\bf p}_{l'}}
\newcommand{\vpc}{{\bf p}_C}
\newcommand{\vp}{{\bf p}}
\newcommand{\vq}{{\bf q}}
\newcommand{\vpf}{{\bf p}_f}
\newcommand{\vpl}{{{\bf p}}_{l}}
\newcommand{\vxl}{{{\bf x}}_l}
\newcommand{\vplz}{{{\bf p}}_l^{(0)}}
\newcommand{\vR}{{{{\bf R}}}}
\newcommand{\dm}{\delta m^2}
\newcommand{\dom}{\delta \omega}
\newcommand{\om}{\omega}
\newcommand{\vpa}{{\bf p}_A}
\newcommand{\vk}{{\bf k}}
\newcommand{\vP}{{{{\bf P}}}}
\newcommand{\vva}{{\bf v}_a}
\newcommand{\vv}{\bf v}
\newcommand{\x}{{\bf x}}
\newcommand{\p}{{\bf p}}
\newcommand{\q}{{\bf q}}
\newcommand{\n}{{\bf n}}
\newcommand{\e}{{\bf e}}
\maketitle

\begin{abstract}

We consider a thought experiment, in which a 
neutrino is produced by an electron on a nucleus in a crystal.
The wave function of the oscillating neutrino is calculated assuming
that the electron is described by a wave packet. If the 
electron is relativistic and the spatial size of its wave packet 
is much larger than the size of the crystal cell, then the wave 
packet of the produced neutrino has essentially the same size  
as the wave packet of the electron.  
We investigate the suppression of neutrino
oscillations at large distances caused by two mechanisms:
1) spatial separation of wave packets  corresponding
to different neutrino masses; 2) neutrino energy
dispersion for given neutrino mass eigenstates. We resolve 
contributions of these two mechanisms.

\end{abstract}

\section{Introduction.}

There are two different approaches to neutrino oscillations in the
literature: one of them deals with the wave function of free
neutrinos, while the other considers the propagator of virtual
neutrinos. The latter approach was analyzed in the papers
\cite{oslica}, \cite{kmos} for a thought experiment,
where a neutrino was considered to be
produced by an electron on the target nucleus $A$, and captured by
the nucleus $B$ in the detector. The process was described as a
two-stage Feynman amplitude, $e + A \to C + \nu $, $\nu + B \to D
+ l$ with a virtual neutrino, its Green function connecting the
production and detection points: $x_A =(t_A, \vx_A)$ and  $x_B
=(t_B, \vx_B)$. As it is well known, at a large distance $r_{AB} =
|\vx_B - \vx_A |$ from the production point virtual particles
become effectively real and one may speak about their wave
function, in particular, about the neutrino wave function, $\psi_\nu
(x)$. Though this statement is well known and practically evident,
an explicit expression for $\psi_\nu$ can be instructive for the
description of the effect of oscillation suppression in a thought
experiment.

There are two mechanisms of erasing oscillations. The first one
is spatial separation of neutrino wave packets of different
mass eigenstates (see refs.\cite{Nussinov} --
\cite{ad-erice}). The second mechanism is caused by neutrino
energy dispersion (see e.g. ref.\cite{PDG}). 
The amplitude of the two-stage process $e + A \to C + \nu_i $ and
$\nu_i + B \to D + l$ with a virtual neutrino of given mass $m_i$
is determined by the standard rules of quantum field theory 
\cite{oslica}: 
\be 
A_i= \int d^4x_1 d^4x_2 \psi^*_l(x_2) \psi_D^* (x_2) \psi_B (x_2) 
G_i(x_2-x_1) \psi_C^* (x_1) \psi_e (x_1) \psi_A(x_1), 
\label{a-i} 
\ee
where $x_{1,2}$ are 4-dimensional coordinates, 
$x_k= (t_k,{\bf x}_k)$, $G_i(x_2-x_1)$ is the Green function 
of the $i$-th neutrino mass
eigenstate, and non-essential spin factors are neglected.
The term $P_{ij} = A_i~A^*_j$ in the probability
was integrated over the phase space of the final particles
in ref.\cite{oslica}. 
After integration the interference term 
 with $i \neq j$ vanishes at large 
spatial separation $| {\bf x}_A - {\bf x}_B| $.

In this approach, however, we were unable to resolve the 
contributions of the two mechanisms. 
In this note we try to separate the "Siamese twins".
For this purpose we consider only the first stage of the
process, $e + A \to C + \nu$, with a free neutrino.
We consider the case when the spatial size of the electron 
wave packet is much larger than the size of the crystal cell,
which determines the localization of the nucleus.
The opposite case will be described elsewhere.

\section{ Wave function {\it vs} amplitude}

The non-normalized wave function of a neutrino produced  
in the reaction $e+A \rar C + \nu$ is
$$\psi_{\nu}=\sum_i U_{ei}\psi_i |\nu_i \rangle, $$
\begin{equation} \psi_i
(t,\x) = \int d^4x_1 G_i (x-x_1) \psi_C^* (x_1) \psi_e (x_1)
\psi_A(x_1), \label{psi-i}
\end{equation}
where $x\equiv(t,\x)$ is the space-time coordinate and 
$|\nu_i \rangle$ is the $i-$th neutrino mass eigenstate. 

In this equation the product 
$\psi_C^* (x_1) \psi_e (x_1) \psi_A(x_1)$ serves as a 
local source of neutrinos.  To calculate the amplitude 
of neutrino interaction with the nucleus $B$ one 
would evidently substitute this expression for $\psi_i$ 
into the integral over $x$ with the wave functions of 
other particles participating in the reaction, according to 
eq.(\ref{a-i}). This naturally agrees with the general 
prescription.

Like in our recent paper \cite{oslica} we consider the initial
nucleus bound in a crystal and describe it by a stationary wave
function localized near the point $\x=0$ :

\begin{equation}
\Psi_A (x) = F_A ({\bf x}) ~ e^{-it E_A}~, \label{Psia}
\end{equation}
where $E_A$ is the energy of the nucleus.
The Fourier transform of $F_A({\bf x})$, which
is required in what follows, is

\begin{equation}
K_A({\bf q}_A) = \int d{\bf x} F_A ({\bf x}) e^{-i{\bf q}_A {\bf
x}}~. 
\label{K_A}
\end{equation}

By assumption, the nucleus $A$ is at rest and, thus, 
$K_A({\bf q}_A)$ is centered at ${\bf q}_A = 0$ with the 
uncertainty  $\sigma_A\sim a^{-1}$.

The wave function of the incident electron is considered to be
a wave packet:
\begin{equation}
\label{Psie}
\Psi_e (x) = \int d{\bf q}_e K_e ({\bf q}_e -{\bf p}_e) e^{i{\bf
q}_e ({\bf x} - {\bf x}_e) - iE_e(\q_e)t} =
e^{i\p_e(\x-\x_e)-iE_e(\p_e) t}~F_e \left( \x- \e v_e t \right).
\end{equation}
Here $\e\equiv{\p_e} / {p_e}$, $p_e \equiv |\p_e|$,
$E_e({\bf q}_e)\equiv \sqrt{{\bf
q}_e^2 +m_e^2}$, the Fourier amplitude $K_e({\bf q}_e -{\bf p}_e)$
is centered  near ${\bf q}_e ={\bf p}_e$ with the 
uncertainty $\sigma_e$, the center of the packet envelope $F_e\left(
\x- \e v_e t \right)$ is at the point $\x_e$ at the moment $t=0$,
and the electron group velocity $v_e$ is defined as
\begin{equation}\label{electron velocity}
v_e\equiv\frac{\partial E_e(\q_e)}{\partial
q_e}\mid_{\q_e=\p_e}=\frac{p_e}{\sqrt{p^2_e+m^2_e}}\simeq 1 -
\frac{m^2_e}{2 p^2_e}.
\end{equation}
In what follows we assume the electron to hit the nucleus 
$A$ at $t=0$ and the collision to be central, i.e. $\vx_e=0$.

The recoil nucleus $C$ is described by the plane wave
\begin{equation}
\Psi_C^* (x) = e^{ i t E_C - i{\bf p}_C {\bf x} },
 \label{Psic}
\end{equation}
unless its momentum is comparable with $\sigma_A$, which is
an extremely rare case.

For the Green function the following expression can be derived
\begin{equation}
G_i (t,\x) = -\frac{1}{4\pi |\x|} \int_{-\infty}^\infty d\om
e^{-i\om t + i \sqrt{\om^2 - m_i^2}\, |\x|} .\label{g}
\end{equation}

Let us now substitute  expressions (\ref{Psia}), 
(\ref{Psie}), (\ref{Psic}) and (\ref{g})
into eq.(\ref{psi-i}) and perform trivial integration
over $x_1$:
$$ \psi_i (t,\x) = \frac{1}{r} \int d
\omega d \q_e K_e \left(\q_e-\p_e \right) d \q_A K_A
\left(\q_A\right) $$
\begin{equation}\label{integrated over dx_1}
 \delta (E_e(\q_e) + E_A - E_C - \omega) 
 \delta (\q_e + \q_A - \p_C - \vk_i)  
\exp ({ik_i r -i \omega t })~.
\end{equation}
Here $r\equiv|\x|$, $\vk_i\equiv \n k_i$, $\n\equiv \x/r$,
$k_i\equiv \sqrt{\omega^2-m_i^2}$ and the expansion

\begin{equation}\label{expansion}
|\x-\x_1| \approx r - \n\x_1,
\end{equation}
is performed 
assuming $r$ to be much larger than the interaction region.  In
eq.(\ref{integrated over dx_1}) and in what follows we omit some
non-essential numerical factors.
We integrate eq.(\ref{integrated over dx_1}) over $\q_A$ and
$\omega$  and obtain
\begin{equation}\label{psi-2}
\psi_i (x) = \frac{1}{r} \int d \q_e K_e \left(\q_e - \p_e\right)
K_A \left(-\q_e+\p_C+\vk_i\right)  \exp \left[ {ik_i r -i \om t } \right],
\end{equation}
where now $\omega(\q_e)\equiv E_e(\q_e) + E_A - E_C$.

While proceeding with the calculations we bear in mind the range
and  hierarchy of the quantities involved:
\begin{equation}\label{hierarchy}
m_i \ll \sigma_e \ll \sigma_A \ll m_e \ll p_e \ll
M_A,M_C.
\end{equation}
In particular, we consider the case when $K_A(\q_A)$ is wide in comparison with
$K_e(\q_e)$:

\begin{equation}\label{condition on sigmas}
\sigma_e\ll\sigma_A.
\end{equation}
 This allows to set $K_A \left(-\q_e+\p_C+\vk_i\right)=K_A
\left(-\p_e+\p_C+\vk^0_i\right)$,
$\vk^0_i\equiv\vk_i\mid_{\q_e=\p_e}$ throughout the  
essential range of
integration over $\q_e$. Assuming that the momentum distribution
of the electron is sufficiently narrow we may expand the integrand
in terms of $\q_e $ near the central electron momentum $\p_e$:
$$\om(\q_e)=\om^0+v_e \e (\q_e-\p_e),$$
\begin{equation}
\label{k i expansion} k_i(\q_e)=
k^0_i+\frac{v_e}{v_i}\e(\q_e-\p_e),
\end{equation}
where  we introduce the neutrino group velocities analogously 
to the electron one:
\begin{equation}\label{neutrino velocity}
 v_i\equiv\left(\frac{\partial\om}{\partial
k_i}\right)^0= \frac{\sqrt{(\om^0)^2-m_i^2}} {\om^0}
\simeq 1-\frac{m_i^2}{2(\om^0)^2};
\end{equation}
the upper index "$0$" means that the corresponding quantities are
calculated at $\q_e = \p_e$.

Taking into account eqs.(\ref{Psie}), (\ref{psi-2}) and (\ref{k i
expansion}) we obtain the following simple expression for the wave
packet of the produced neutrino through the envelope of the wave
packet of the incoming electron in coordinate space, $F_e$:

\be 
\psi_i(t,r,\n) = \frac{e^{ik^0_i r- i\om^0
t}}{r}K_A(-\p_e+\p_C+k^0_i\n) 
F_e \left(\frac{v_e}{v_i}(r-v_i t)\e\right)~. 
\label{psi-fin}
\ee

The factor $v_e/v_i\simeq 1$ makes the neutrino wave packet a little
bit wider then the electron one. It is not essential for our
purposes and will be omitted in what follows.

Equation (\ref{psi-fin}), which is one of the main results of this
paper, is quite natural. If there were a long wave packet of the
incoming  electron, it would create a packet of neutrinos with a
similar length. A good analogy is the scattering of a sound wave
on a target which creates another sound wave. The duration and,
correspondingly, the spatial length of the produced wave packet
should be equal to the duration and size of the original one.

Strictly speaking, one has to calculate the amplitude
(\ref{a-i}) to determine the probability of the oscillating behavior
of neutrinos. However, one may rely on a simplified approach based
on the interpretation of the absolute
value squared of the neutrino wave function as the probability 
density for the particle to be found at a
spatial point $\x$ at a given time $t$. Such an approach is valid, if we
deal with wave packets,  the longitudinal size of which 
is much larger then their  wave length, and their transversal size
is much larger than their Compton wave length
(see, e.g.,~\cite{Bjorken-Drell}, \cite{Weinberg}). If we
neither register the nucleus $C$ nor measure the time of neutrino
detection, we are interested in the detection probability of 
the neutrino $\nu_l$ at point $\x$:
\begin{equation}
\label{Probability}
P_{\nu_e \rightarrow \nu_l}(\x)=
\int d\p_C dt \left|\sum_i U_{li}^* U_{ei}
\psi_i(t,\x) \right |^2.
\end{equation}

For simplicity, in what follows we assume $i=1,2$, 
$~l=e,\mu$, $~U_{e1}=\cos \theta,~U_{e2}=\sin \theta$. For the
$\nu_{\mu}$ production probability we obtain

$$P_{\nu_e \rightarrow \nu_{\mu}}(\x)=\frac{f}{2r^2}(\sin
2\theta)^2\int d\p_C K^2_A(-\p_e+\p_C+k^0_i\n)$$
\begin{equation}\label{nu-mu appearance probability}
\left[1-\frac{1}{f}\cos \left(\frac{r\delta
m^2}{2\om^0}\right)\int dt 
F_e \left((r-v_1t)\e\right)F_e\left((r-v_2t)\e\right)\right],
\end{equation}
where $f\equiv\int dt F^2_e\left((r-t)\e\right)$, 
$\delta m^2 \equiv m_2^2 -m_1^2$.

The term proportional to $\cos (r\delta m^2 / 2\om^0)$ 
in eq.(\ref{nu-mu appearance probability})
describes oscillations. It vanishes at large distances for two
different reasons.\\
{\bf 1. Packet separation.}\\
When $(v_2-v_1)r>{1} / {\sigma_e}$, that is
\begin{equation}
\label{condition for packet separation}
r>L_{osc}\frac{p_e}{\sigma_e},~~~L_{osc}\equiv \frac{2p_e}{\delta
m^2 }
\end{equation}
 the product
$F_e(r-v_1t)F_e(r-v_2t)$ is nearly  zero for every $t$, and
the oscillating term vanishes even {\bf before} integration over
$\p_C$. \\
{\bf 2. Neutrino energy dispersion.}\\
Note that  $\om^0$ depends on $\p_C$:
\begin{equation}\label{omega 0}
 \om^0\simeq E_e(\p_e)+E_A-M_C-\frac{p^2_C}{2M_C},
\end{equation}
where $M$ denotes the $C$ nucleus  mass. Owing to the factor
$K^2_A(-\p_e+\p_C+k^0_i\n)$ in eq.(\ref{nu-mu appearance
probability}), the effective size of the region of integration over
$\p_C$ is of the order of $\sigma_A$. Thus $\omega^0$ varies near its
central value (which is roughly equal to $p_e$), and the variation
equals  ${p^0_C\sigma_A}/{M_C}$. Here, the central value $\p^0_C$
is determined by the equation
\begin{equation}\label{equation for p^0_C}
\p^0_C=\p_e-\left(p_e+\frac{m^2_e}{2p_e}+
E_A-M_C-\frac{(p^0_C)^2}{2M_C}\right)\n.
\end{equation} 
If the angle between
vectors $\e$ and $\n$ is sufficiently large ($|\e-\n|p_e\gg
~ M_C-E_A,~{p^2_e} / {2M_C}$), then
\begin{equation}
\label{final p^0_C}
p^0_C\simeq p_e|\e-\n|.
\end{equation}
This is the case  considered below. The variation
of $\om^0$  equals  ${p_e|\e-\n|\sigma_A}/{M_C}$. Thus integration
of $\cos \left({r\delta m^2} / {2\om^0}\right)$ over $\p_C$ in
eq.(\ref{nu-mu appearance probability}) leads to the vanishing
result if
\begin{equation}\label{condition for the dispertion suppression}
r>L_{osc}\frac{M_C }{|\e-\n|\sigma_A}.
\end{equation}

We see that there are two competitive mechanisms for suppression
of  neutrino oscillations.  If
\begin{equation}\label{very small sigma_e}
\sigma_e < \sigma_A \frac{p_e |\e-\n|}{M_C},
\end{equation}
then the energy dispersion mechanism dominates, and
\begin{equation}\label{energy dispersion suppression length}
L_{sup}=L_{osc}\frac{M_C }{|\e-\n|\sigma_A}.
\end{equation}

If
\begin{equation}\label{not very small sigma_e}
\sigma_A \frac{p_e |\e-\n|}{M_C}<\sigma_e \ll \sigma_A,
\end{equation}
then the packet separation works, and
\begin{equation}\label{packet separation suppression length}
L_{sup}=L_{osc}\frac{p_e }{\sigma_e}.
\end{equation}
These results coincide with those obtained in ref.\cite{oslica}.

One more comment is worth making at this stage. As we said above, the
interference disappears when two neutrino wave packets $\psi_i$ and
$\psi_j$, eq.(\ref{psi-fin}), cease to overlap. At first sight this 
statement is at
odds with the expression for the production amplitude (\ref{a-i})
of lepton
$l$ on the nucleus $B$. Indeed, the product of amplitudes 
$A_i$ and $A_j^*$, which enters the probability of the process
(see eqs.(27)-(29) of ref.\cite{oslica}), does
not vanish even when the product $\psi_i (x)\psi_j^*(x)$ vanishes, because
the amplitude contains an integral over $x$ and the product of integrals
never vanishes. However, one can check that after integration over the phase
space of the final particles the product of the integrals vanishes exactly
when the neutrino wave packets no longer overlap. Such integration over 
the phase space makes the result effectively local.

This conclusion is intuitively clear for the following reasons. The product
of the amplitudes prior to integration over the phase space describes the
production probability of a plane wave final state, because the final
states are taken as momentum eigenfunctions. It is evident that such a
probability never vanishes even in the case of infinite separation of neutrino
wave packets. It is essentially the same as the excitation of a resonator
by two wave packets. A resonator with a very large $Q$-factor would stop to
oscillate only after a very long time. So, if such a resonator is hit by one
wave packet and after a while by another delayed wave packet, the
interference between the packets would  still
be observed by such a resonator
because it keeps oscillating long after the fist packet has gone while the
second has just arrived.

\section{Conclusions}

In this note we have considered neutrinos produced in the 
reaction $e+A \to C+\nu$. For the case of a large
size of the electron wave packet ($\sigma_e \ll \sigma_A$)
we have calculated the neutrino wave packet 
(see eq.(\ref{psi-fin})). Its size coincides with that
of the incident electron wave packet.

We have demonstrated that in the case of
$\sigma_e < p_e|{\bf e} - {\bf n}| \sigma_A /M_C$
the suppression of neutrino oscillations at large
distances is due to the neutrino energy dispersion,
while in the case of
$p_e|{\bf e} - {\bf n}|\sigma_A /M_C < \sigma_e \ll \sigma_A$ 
it occurs because of the neutrino packet separation. 
The corresponding suppression lengths are given by 
eqs.(\ref{energy dispersion suppression length})
and (\ref{packet separation suppression length}).
It is evident that for terrestrial conditions such lengths
are unrealistic.

\section{Acknowledgments}

We are grateful to G.B. Pontecorvo and M.V. Rotaev for valuable discussions.

This work was partially supported by Dynasty Foundation, by the
grant SH-2328.2003.2 and by the RFBR grant 05-02-17471.

\end{document}